\documentclass[aps,pre,twocolumn,superscriptaddress]{revtex4}

\usepackage{latexsym}
\usepackage{amstext,amssymb,amsfonts}
\usepackage{epsfig}
\usepackage{graphics}

\begin{document}

\title{Energy-landscape network approach to the glass transition}
\author{Shai Carmi}
\affiliation{Minerva Center \& Department of Physics, Bar-Ilan
University, Ramat Gan 52900, Israel} \affiliation{Center for Polymer
Studies, Boston University, Boston, MA 02215 USA}
\author{Shlomo Havlin}
\affiliation{Minerva Center \& Department of Physics, Bar-Ilan
University, Ramat Gan 52900, Israel}
\author{Chaoming Song}
\affiliation{Levich Institute and Physics Department, City College
of New York, New York, New York 10031, USA}
\author{Kun Wang}
\affiliation{Levich Institute and Physics Department, City College
of New York, New York, New York 10031, USA}
\author{Hernan A. Makse}
\affiliation{Levich Institute and Physics Department, City College
of New York, New York, New York 10031, USA}

\date{\today}

\pacs{64.70.Q-,64.60.aq,64.60.ah,89.75.Hc,89.75.Fb}

\begin{abstract}

We study the energy-landscape network of Lennard-Jones clusters as a
model of a glass forming system. We find the stable basins and the
first order saddles connecting them, and identify them with the
network nodes and links, respectively. We analyze the network
properties and model the system's evolution. Using the model, we
explore the system's response to varying cooling rates, and
reproduce many of the glass transition properties. We also find that
the static network structure gives rise to a critical temperature
where a percolation transition breaks down the space of
configurations into disconnected components. Finally, we discuss the
possibility of studying the system mathematically with a trap-model
generalized to networks.
\end{abstract}

\maketitle

\section{Introduction}

In recent years much effort was devoted to the understanding of
supercooled liquids and structural glasses, and, in particular, the
structural arrest taking place at the glass transition temperature
$T_g$ \cite{SupercooledReview,kob}. The numerical investigation of
the dynamics of supercooled liquids and glasses is very hard due to
the presence, approaching $T_g$, of this very slow dynamics
\cite{Sastry}. An appealing approach for understanding this complex
dynamics is to study the properties of the system's ``energy
landscape'': the dynamics of the system is viewed as the motion of
the ``state point'', described by the $3n$-coordinates of all
particles in the multi-dimensional configuration space, or
landscape, of the potential energy of the system ($n$ is the number
of particles). The landscape may be partitioned into ``basins of
attraction'', such that local minimization of the potential energy
maps any point in a basin to the same minimum. In recent years it
has been shown that the topological details of the basins and the
paths connecting them are of great importance in determining the
properties of glassy systems (e.g.,
\cite{Parisi98,AngelaniPRE,Angelani,Heuer,Grigera,Doliwa1}).

The representation of the landscape by its basins leads to a further
simplified view of the energy landscape as a network, where the
nodes are the basins and the links are the saddles connecting them.
The energy-landscape network of a Lennard-Jones (LJ) system has been
mapped, and some of its properties were extracted \cite{Doye,Doye2}
(for energy-landscape networks in proteins and spin systems see
\cite{Caflisch,gradient_protein,SpinNetwork}). However, the
influence of the topology of the network on the dynamics of the
glass transition was never studied. Here, we characterize the
networks of mono- and bi-disperse LJ systems obtained by
minimization of the potential energy, and use a dynamical model to
study properties such as response to cooling. The integration of the
landscape picture with network theory provides an interpretation of
the different critical temperatures of the glass transition $T_0$
(Vogel–-Tammann–-Fulcher temperature \cite{SupercooledReview,kob})
and $T_g$, as well as identification of a new critical temperature
$T_p$ where a second order phase transition separates a phase where
a finite fraction of the configurations are available, and a phase
with a vanishing number of accessible states.

A network model for the glass transition was introduced in
\cite{Parisi98,AngelaniPRE}. Here, we take advantage of more
sophisticated network analysis tools such as percolation theory. In
particular, our approach takes into account the heterogeneity in the
number of connections of each basin (i.e., its degree $k$), which
was recently shown to be ubiquitous in nature and crucial for the
understanding of many networks' properties \cite{BA_review}.

\section{The network's static properties}
\label{static}

We start with a detailed analysis of the static properties of the
energy-landscape network. We focus on isolated small LJ systems of
two types: \emph{(i)} Monodisperse LJ system (MLJ) with $n=12,14$
particles. \emph{(ii)} Binary 80/20 LJ mixture (BLJ) with $n=8+2$
particles. Our network reflects the landscape of potential energy
(not free energy), or in other words, the entropy is not taken into
account, since we assume that all basins are equivalent in terms of
the number of internal states they represent. We note that the sizes
of the systems we study are small compared to other systems in which
molecular dynamics is run \cite{Sastry}. However, this is inevitable
since the number of nodes increases exponentially with the number of
particles and thus larger systems are computationally much harder to
study \cite{Doye}.

To construct the energy-landscape network, we look for basins, the
local minima which form the network's nodes, and their transition
states-- first order saddles which connect two local minima and form
the network's links \cite{Doye}. We use the LBFGS algorithm
\cite{kw1} to find the basins, and the Eigenmode Method \cite{kw2}
to find the saddles. Sometimes more than one first order saddle
connect two linked basins since the landscape is a high dimensional
surface. To simplify the network, we consider only the saddle with
the minimum energy barrier between two linked basins. While the BLJ
system is known to be glassy \cite{Sastry}, to avoid the
crystallization process usually observed in monodisperse systems, we
do not consider the state of lowest energy when setting up the
network \cite{AngelaniPRE}. Thus, the two systems are expected to be
comparable in terms of their glassy behavior.

\begin{figure}[t]
\begin{center}
\epsfig{file=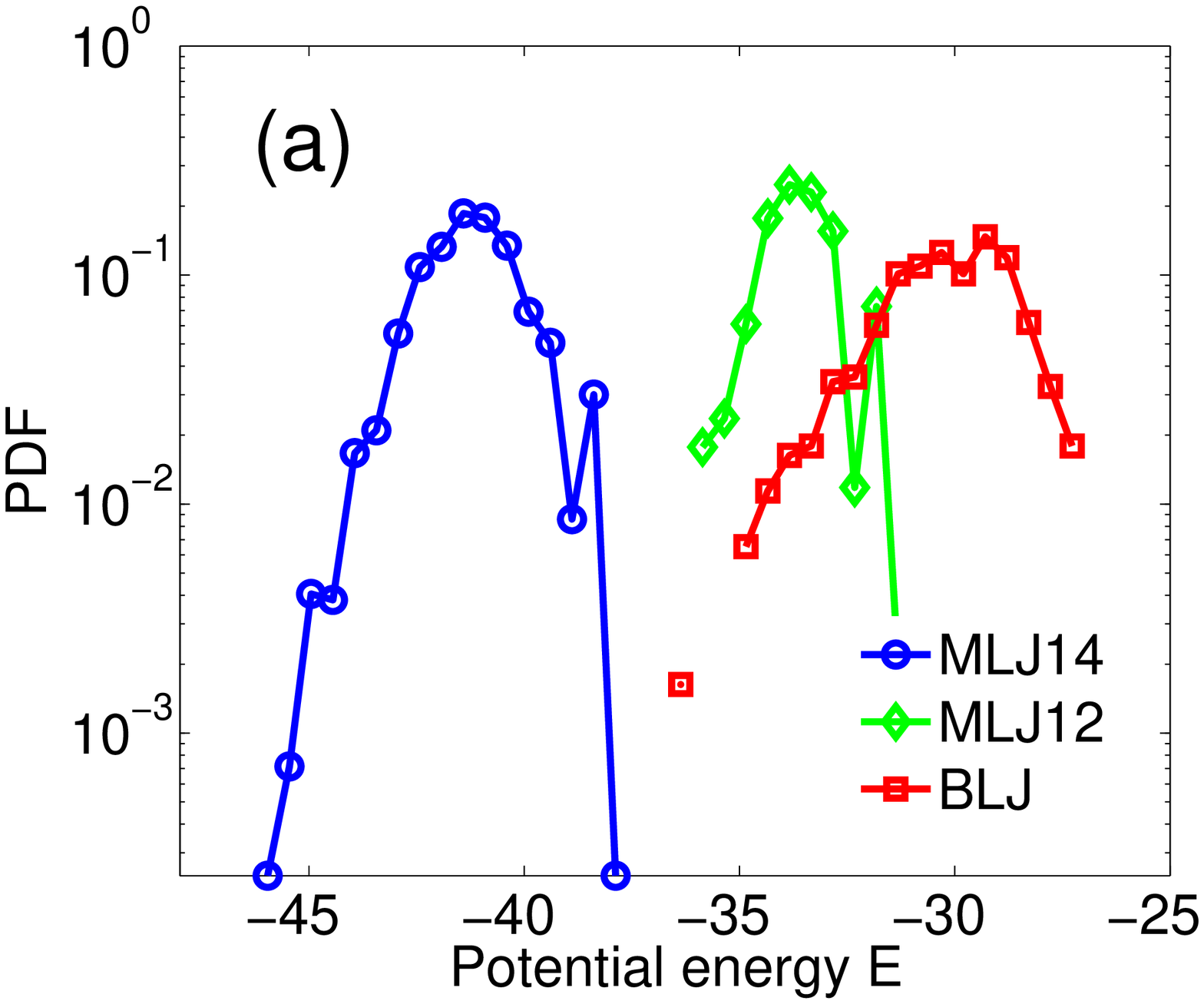,height=3cm,width=4cm}
\epsfig{file=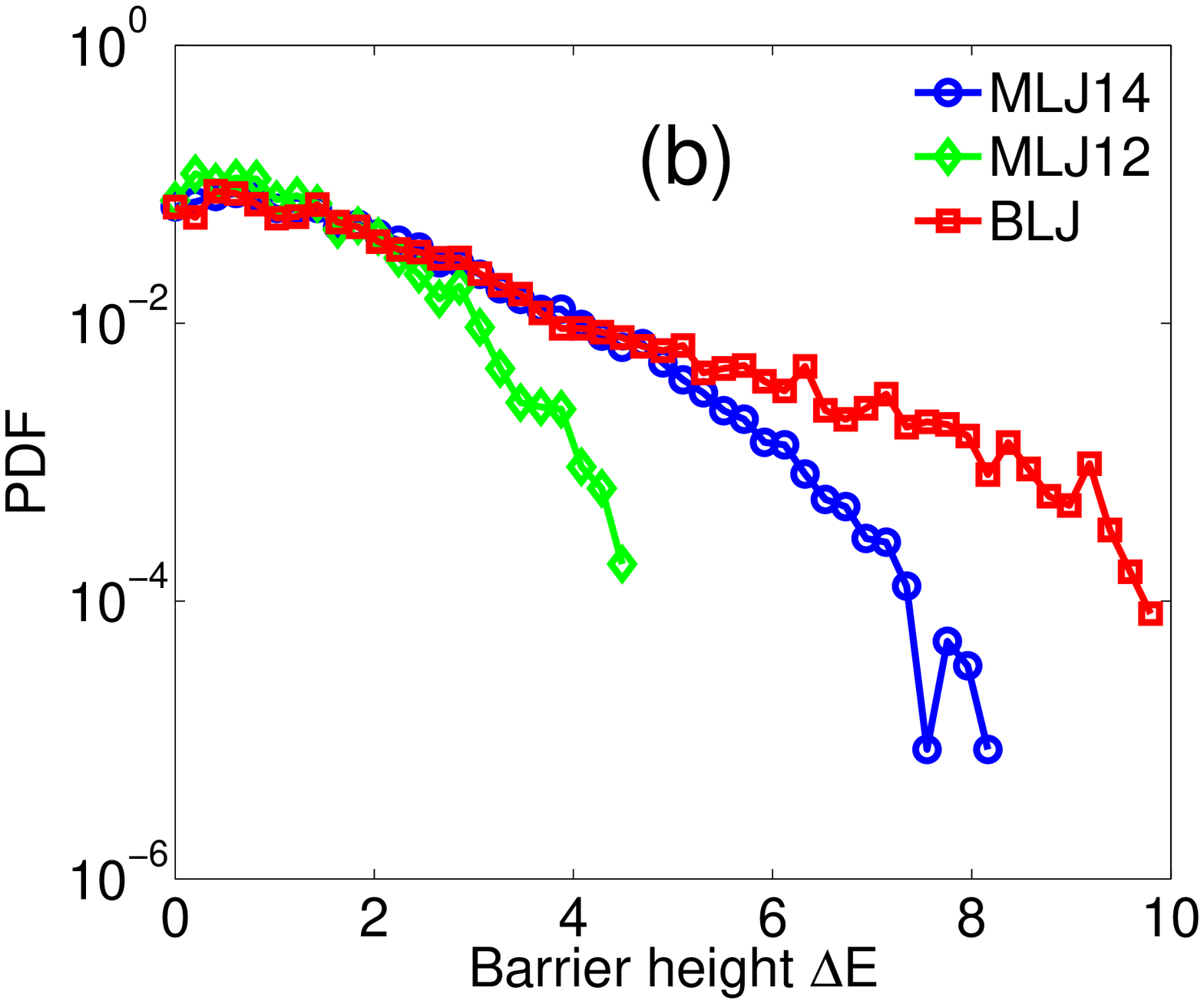,height=3cm,width=4cm}
\epsfig{file=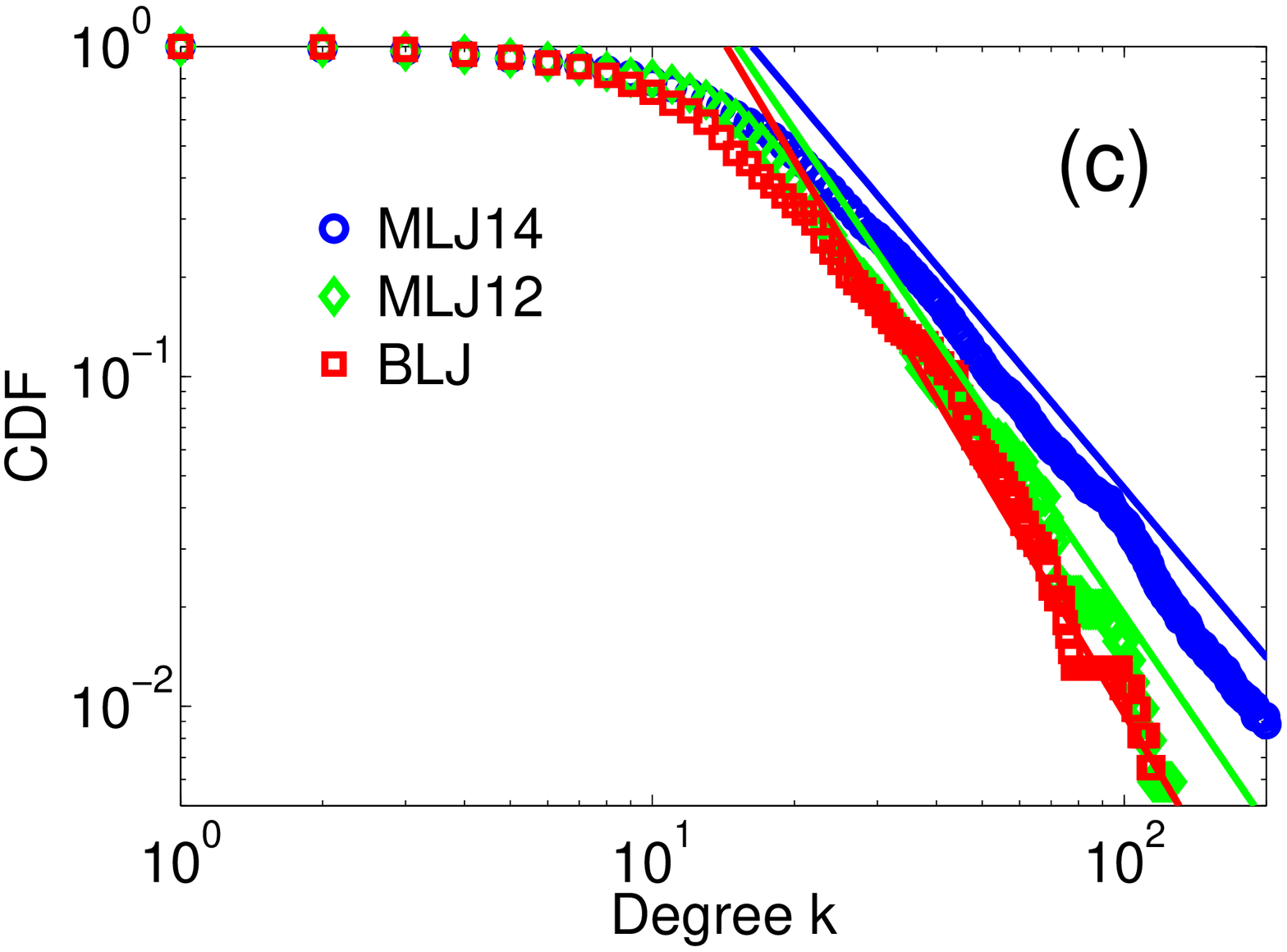,height=3cm,width=4cm}
\epsfig{file=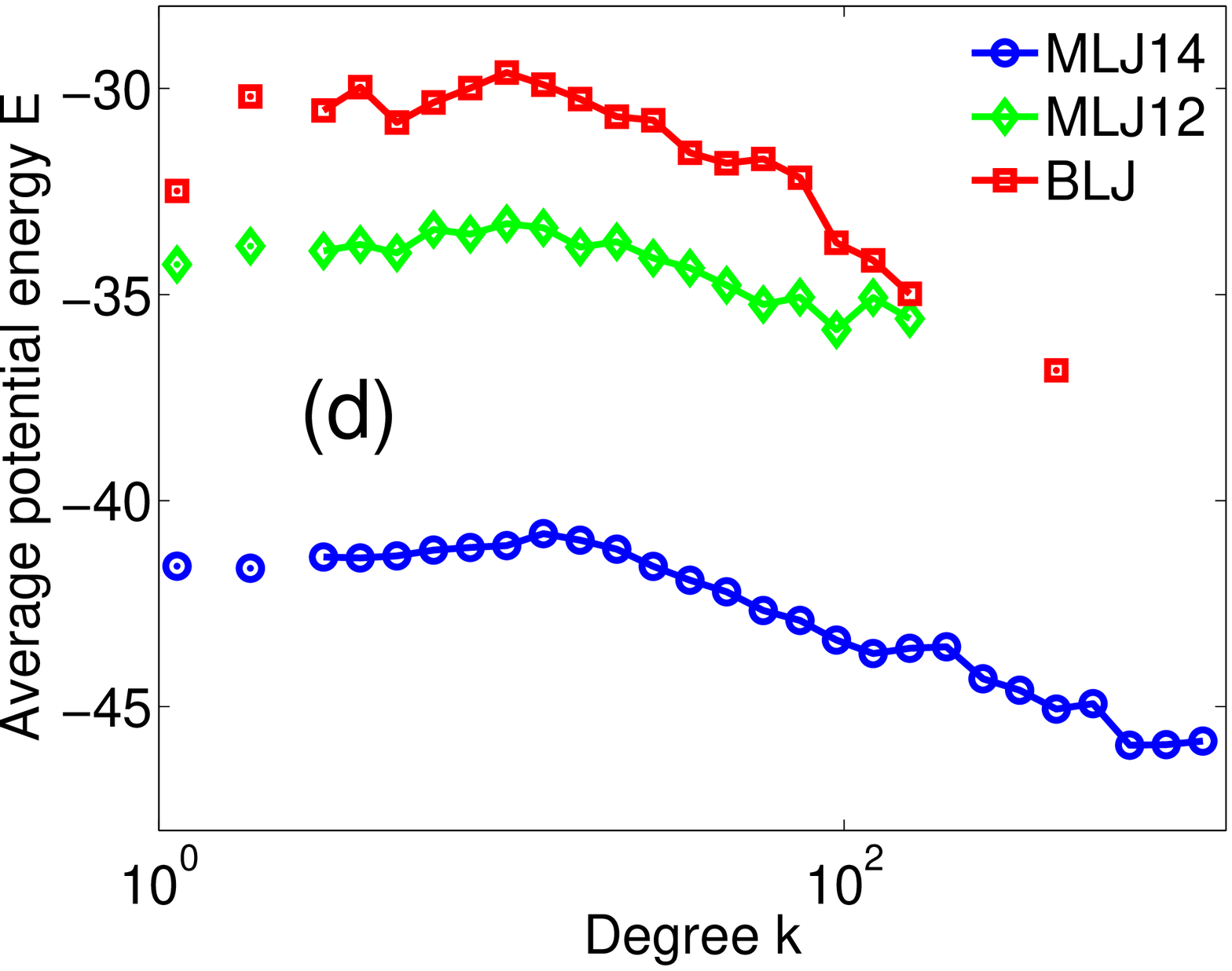,height=3cm,width=4cm}
\epsfig{file=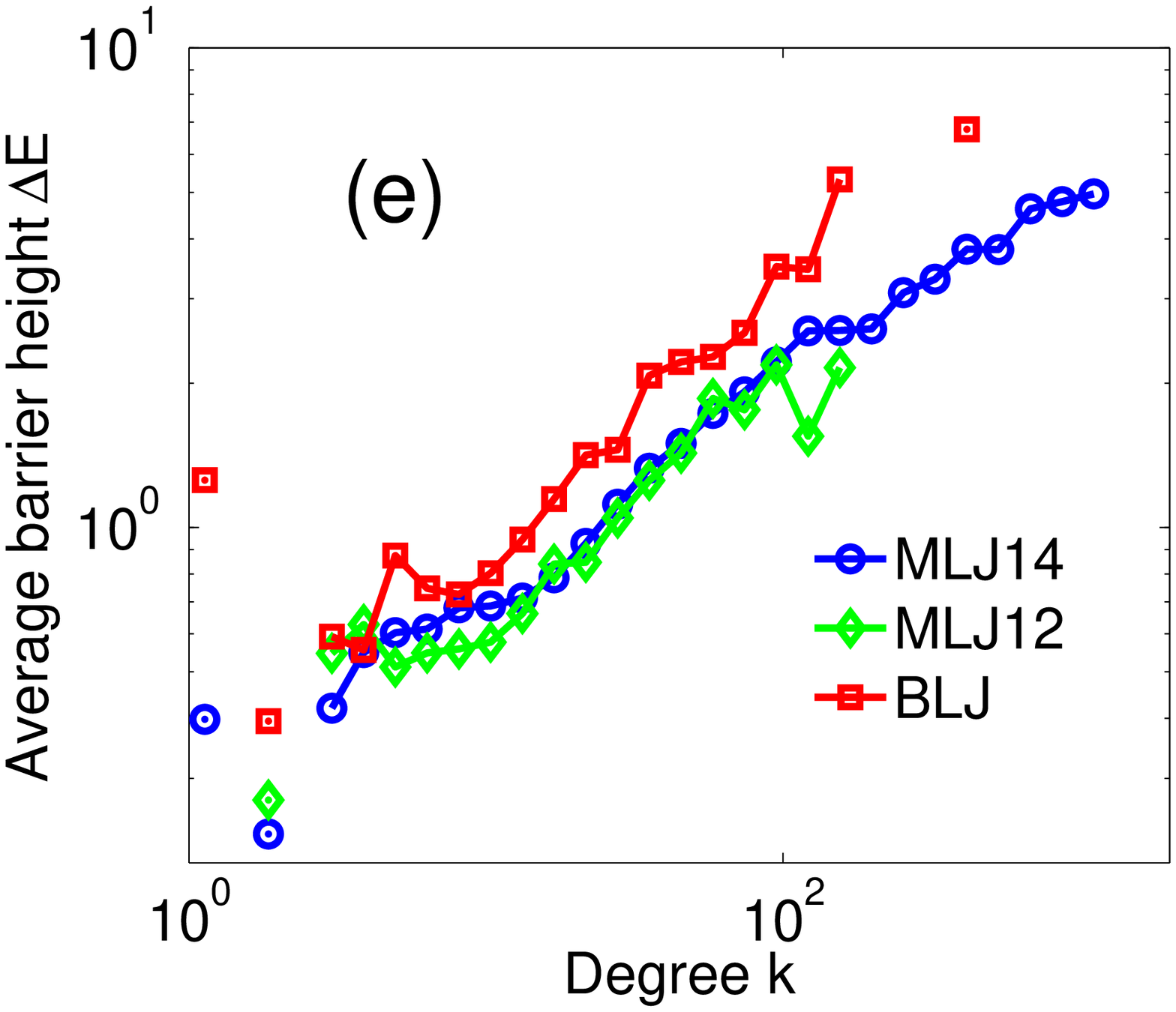,height=3cm,width=4cm}
\epsfig{file=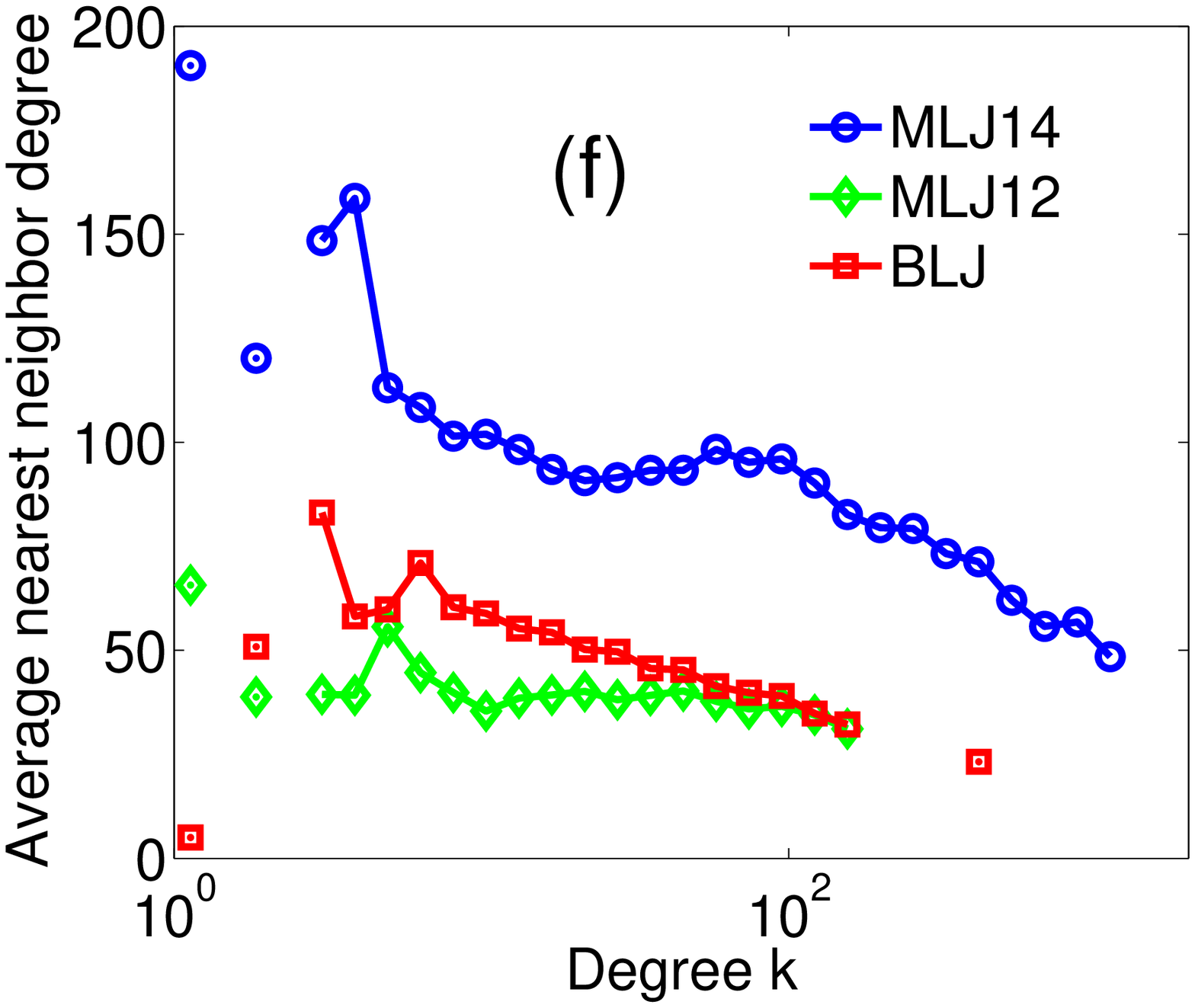,height=3cm,width=4cm}
\end{center} \caption{Properties of the Lennard-Jones energy-landscape network.
Shown are results for MLJ$_{14}$, MLJ$_{12}$, and BLJ (see text). In
all figures the data was binned and the average is plotted. (a)
Distribution of potential energies $E$ of the nodes. (b)
Distribution of the heights of the energy barriers $\Delta E$
associated with the network links. (c) Cumulative distribution of
node degrees $k$. Straight lines represent power-law decays of the
form $P(k)\sim k^{-\gamma}$, with $\gamma=2.7,3.1,3.4$ for
MLJ$_{14}$, MLJ$_{12}$, and BLJ, respectively. (d) The average
potential energy of a node $E$ vs. the degree $k$. (e) The average
energy barrier to escape from a node $\Delta E$ vs. the node degree
$k$. (f) The average degree of node neighbors $\overline{k_{nn}}$
vs. the node's degree $k$.} \label{properties}
\end{figure}

The MLJ$_{14}$ network consists of $N=4,193$ nodes and $M=58,628$
links. The energies at the nodes are distributed approximately
normally with mean $\overline{E}(T\rightarrow \infty)=-41.5$ (Fig.
\ref{properties}(a)). The energy barriers are distributed
approximately exponentially $P(\Delta E) = \frac{1}{\overline{\Delta
E}}e^{-\Delta E/\overline{\Delta E}}$, where $\overline{\Delta
E}=1.64$ is the average energy barrier (Fig. \ref{properties}(b)).
As observed in \cite{Doye}, we confirm that the MLJ$_{14}$ network
is \emph{scale-free} \cite{BA_review}, i.e. the degree distribution
(the probability for a node to have degree $k$) is broad with a tail
decaying as $P(k)\sim k^{-\gamma}$ with $\gamma\approx2.7$ (Fig.
\ref{properties}(c)). The energy of a node decreases with its degree
(Fig. \ref{properties}(d)), meaning that the deepest basins can be
identified with the network \emph{hubs}, and are thus accessible
to/from many other basins \cite{KCM_note}. The average barrier
height increases with the degree of the node: $\overline{\Delta E}
\sim k^{\epsilon}$ with $\epsilon \approx 0.43$ (Fig.
\ref{properties}(e)). The average degree of a node's nearest
neighbors $\overline{k_{nn}}$ slightly decreases with the node's
degree (Fig. \ref{properties}(f)), meaning hubs have many
connections to low degree nodes.
We also studied ``slices'' of the network in the degree (using the
$q$-core method \cite{DIMES}) and energy planes. In both cases, we
found that removing nodes of low degree, or high potential energy,
leaves the network connected.

The MLJ$_{12}$ system is smaller and contains only $N=508$ nodes and
$M=5,407$ links, leading to larger fluctuations in its statistics.
Yet the properties of MLJ$_{12}$ are qualitatively similar to
MLJ$_{14}$. For MLJ$_{12}$ we find $\overline{E}(T\rightarrow
\infty)=-33.87$, $\overline{\Delta E}=1.16$, $\gamma\approx3.1$, and
$\epsilon\approx0.41$ (Fig. \ref{properties}). All the results
reported henceforth as MLJ are for the MLJ$_{14}$ system, unless
explicitly otherwise specified. This picture holds true also for the
BLJ network, with $N=613$ nodes and $M=6,150$ links. In the BLJ
network we obtain $\overline{E}(T\rightarrow \infty)=-30.5$,
$\overline{\Delta E}=1.86$, $\gamma\approx3.4$, and
$\epsilon\approx0.54$ (Fig. \ref{properties}).

\begin{figure}[t]
\begin{center}
\hspace{-0.5cm}
\includegraphics[width=9cm]{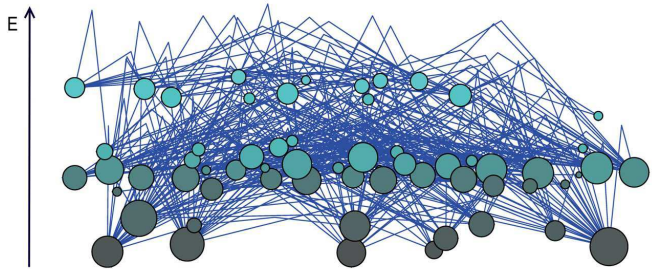} \vspace{-0.5cm}
\end{center}\caption{(Color) MLJ$_{10}$ network schematic.
The vertical axis represents the energy, such that nodes with deeper
energy are lower (and darker) in the schematic. Links' cusps
correspond to the energies of the saddles, and nodes' sizes are
proportional to their degree. It can be seen that highly connected
nodes usually correspond to deeper basins.} \label{picture}
\end{figure}

\section{The network dynamics}

Next we turn to a characterization of the dynamics of the system. We
show that application of simple assumptions about the dynamics
reproduces many features of the glass transition. At high
temperatures, kinetic energy permits access to most states, while
for low temperatures, mutual access among basins becomes subject to
considerable 'activation'. In low temperatures near the transition
(more precisely, below the so-called dynamic glass transition
temperature $T_d$, where an exponential number of meta-stable states
appears \cite{kob}), the dynamics is dominated by rare events of
collective jumps among different stable positions involving many
atoms. Thus, for low temperatures, we neglect the short time
dynamics which is dominated by small vibrations within the basins,
and model the dynamics of the system as activated jumps between
connected states. We assume the transition rate between a pair of
linked states follows Arrhenius law:
\begin{equation}
\label{transition} p_{ij} = \frac{1}{N-1}e^{-\Delta E_{ij}/T},
\end{equation}
where $\Delta E_{ij}$ is the height of the barrier separating $i$
and $j$ (not necessarily equal to $\Delta E_{ji}$) and the $1/(N-1)$
factor guarantees that the rate of leaving $i$, equals to
$\sum_{\textrm{all links \emph{(i,j)}}}p_{ij}$ is less than 1 for
any node. Note there is considerable probability for the system to
remain at the current state. This will turn useful in the
characterization of the dynamical slowdown.

Experiments \cite{SupercooledReview,kob} and molecular dynamics
simulations \cite{Sastry} show that supercooling below the melting
point results in a decrease in the system's energy, up to the
temperature of the glass transition $T_g$. At the transition, the
system becomes frozen in a disordered configuration, and the rate of
change of energy with respect to temperature decreases abruptly (but
continuously) to a value comparable to that of a crystalline solid.
We suggest, that this picture, as well as the identification of the
glass transition temperature $T_g$ can be reproduced using our
simple network dynamics.

\begin{figure}[t]
\begin{center}
\includegraphics[width=4cm]{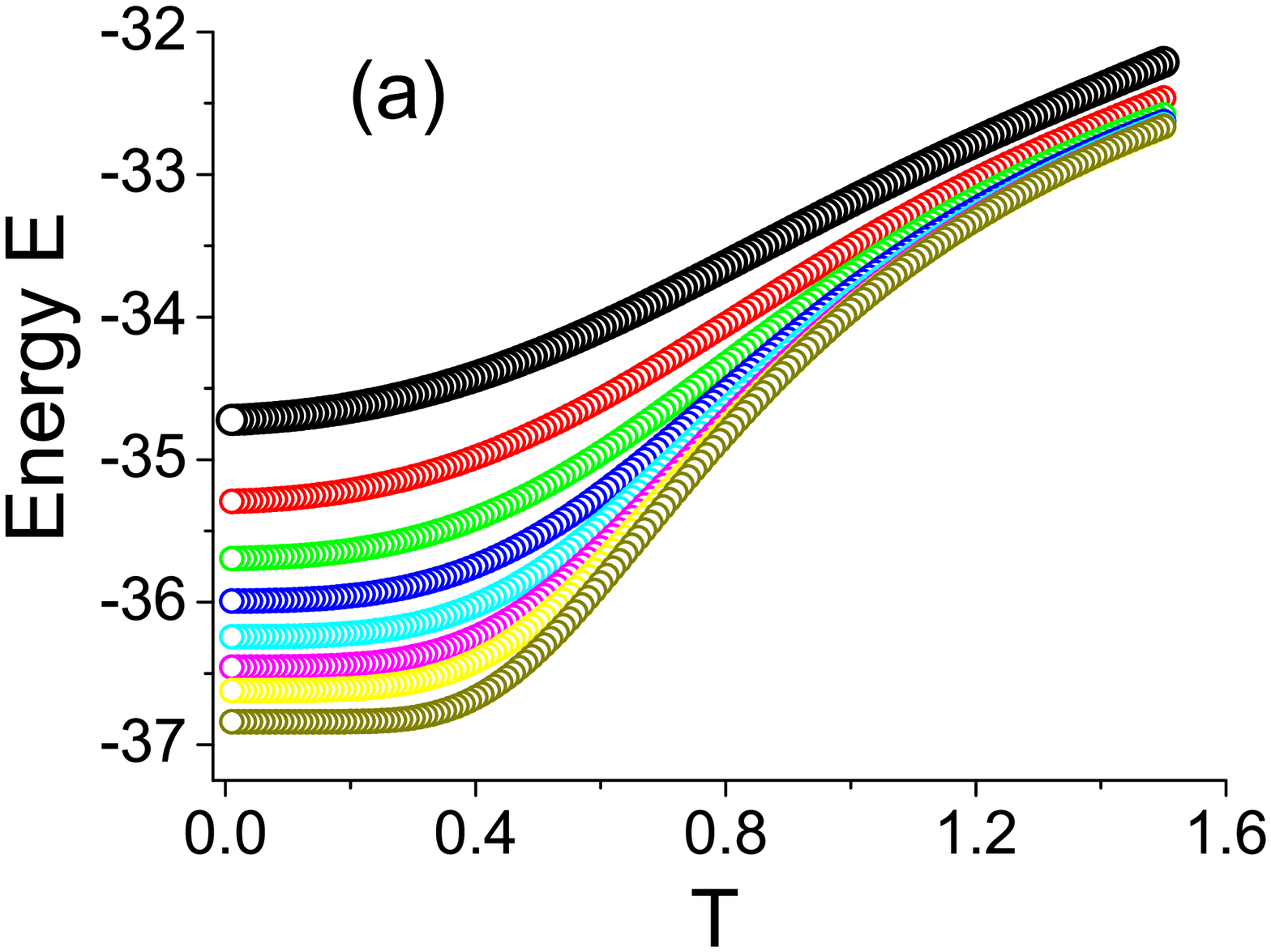}
\includegraphics[width=4cm]{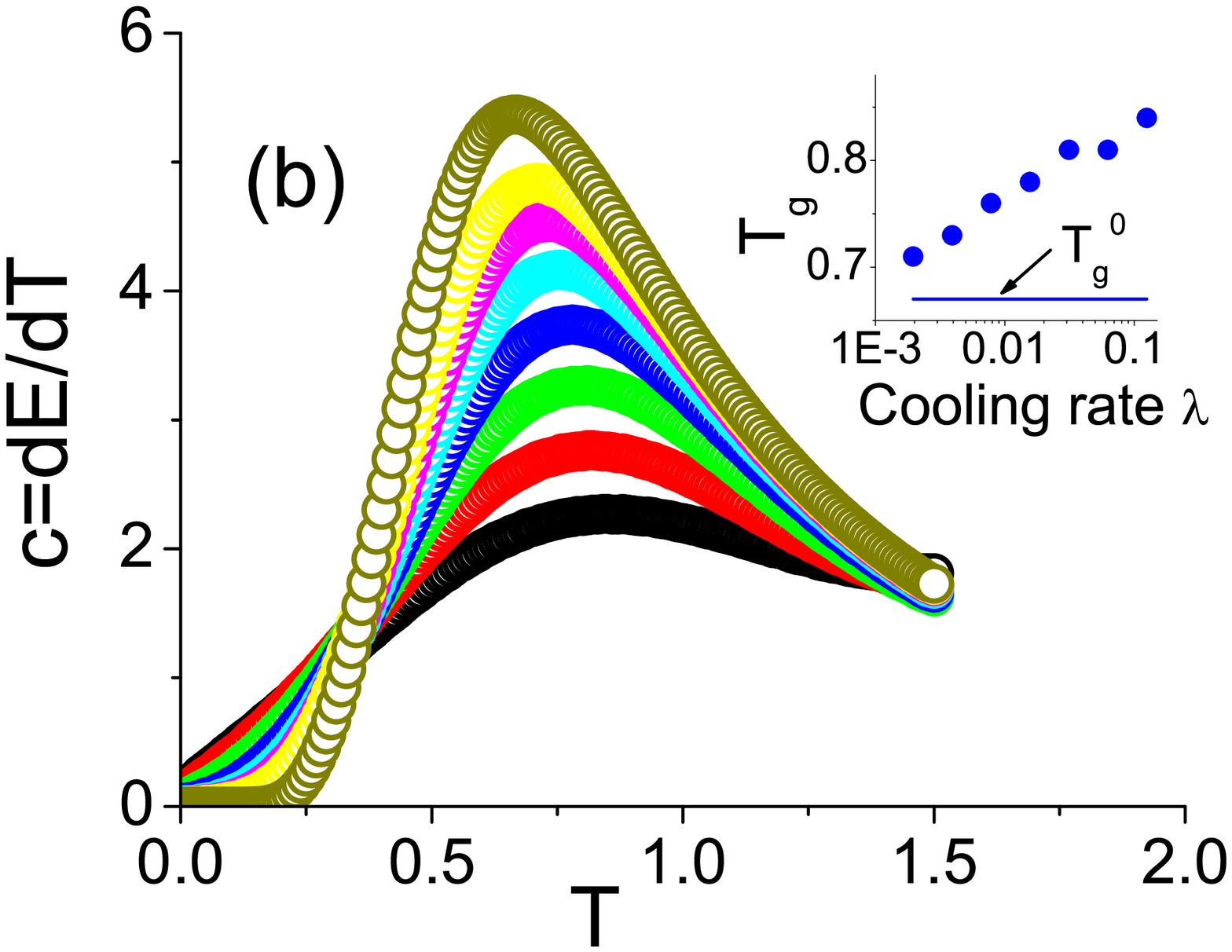}
\vspace{-0.5cm}
\end{center} \caption{(a) Super-cooling in BLJ.
The average energy of the system $\overline{E}$ is plotted vs. the
temperature $T=T_i-\lambda t$. $T_i=2$ and the cooling rates are
(top to bottom): $\lambda=0.01\times\{1/8,1/16,...,1/512,0\}$, where
in each time step we iterate Eq. (\ref{phi}) once. Zero cooling rate
corresponds to the equilibrium Boltzmann distribution. At $t=0$ we
assumed all states are equally probable. Similar results are found
for MLJ (not shown). (b) The heat capacity $c=d\overline{E}/dT$.
$\lambda$ (bottom to top) is same as in (a). Inset: the glass
transition temperature $T_g$ as a function of the cooling rate
$\lambda$. The horizontal line corresponds to $T_g^0$.}
\label{cooling_fig_BLJ}
\end{figure}

$\Phi_i(t)$, the probability of the system to be at state $i$ at
time $t$, evolves according to:
\begin{equation}
\label{phi} \frac{d\Phi_i}{dt} = \frac{1}{N-1}\sum_{\textrm{all
links \emph{(i,j)}}}\Phi_je^{-\Delta E_{ji}/T(t)}-\Phi_ie^{-\Delta
E_{ij}/T(t)}
\end{equation}
We solve this set of equations numerically by iterating Eq.
(\ref{phi}) once in every time step for the MLJ and BLJ networks. We
use different cooling rates $T(t) = T_i-\lambda t$, where $T_i$ is
the initial temperature, and $\lambda$ is the cooling rate. We then
calculate $\overline{E}(T)=\sum_{i}\Phi_i(T(t))E_i$, where $E_i$ is
the energy of node $i$. For infinitely slow cooling, the system can
be assumed to be in equilibrium, such that $\frac{d\Phi_i}{dt}$
vanishes for all $i$. $\overline{E}(T)$ is calculated by setting the
Boltzmann distribution $\Phi_i(T)=e^{-E_i/T}/{\cal Z}$, where ${\cal
Z}=\sum_i e^{-E_i/T}$. The results for BLJ, with
$\lambda=0.01\times\{1/8,1/16,...,1/512,0\}$, are plotted in Fig.
\ref{cooling_fig_BLJ}(a). Indeed we find that our approach
qualitatively reproduces the glass-forming behavior.

We then calculate the heat capacity $c=d\overline{E}/dT$ (Fig.
\ref{cooling_fig_BLJ}(b)). We associate the temperature for which
the heat capacity is maximal with the glass transition temperature
$T_g$. We note that while this association is plausible, it cannot
be made rigorous. As expected
\cite{SupercooledReview,kob,Sastry,old_experiment}, $T_g$ decreases
as the cooling rate becomes slower, approaching its equilibrium
value $T_g^0=0.67(\pm0.01)$ for BLJ (Fig. \ref{cooling_fig_BLJ}(b),
inset). This value of $T_g$ is a little higher than the glass
transition temperature in a large BLJ system ($\approx 0.45$)
\cite{Sastry}. For MLJ, the picture is similar with
$T_g^0=0.47(\pm0.01)$.

\begin{figure}[t]
\begin{center}
\includegraphics[width=4cm]{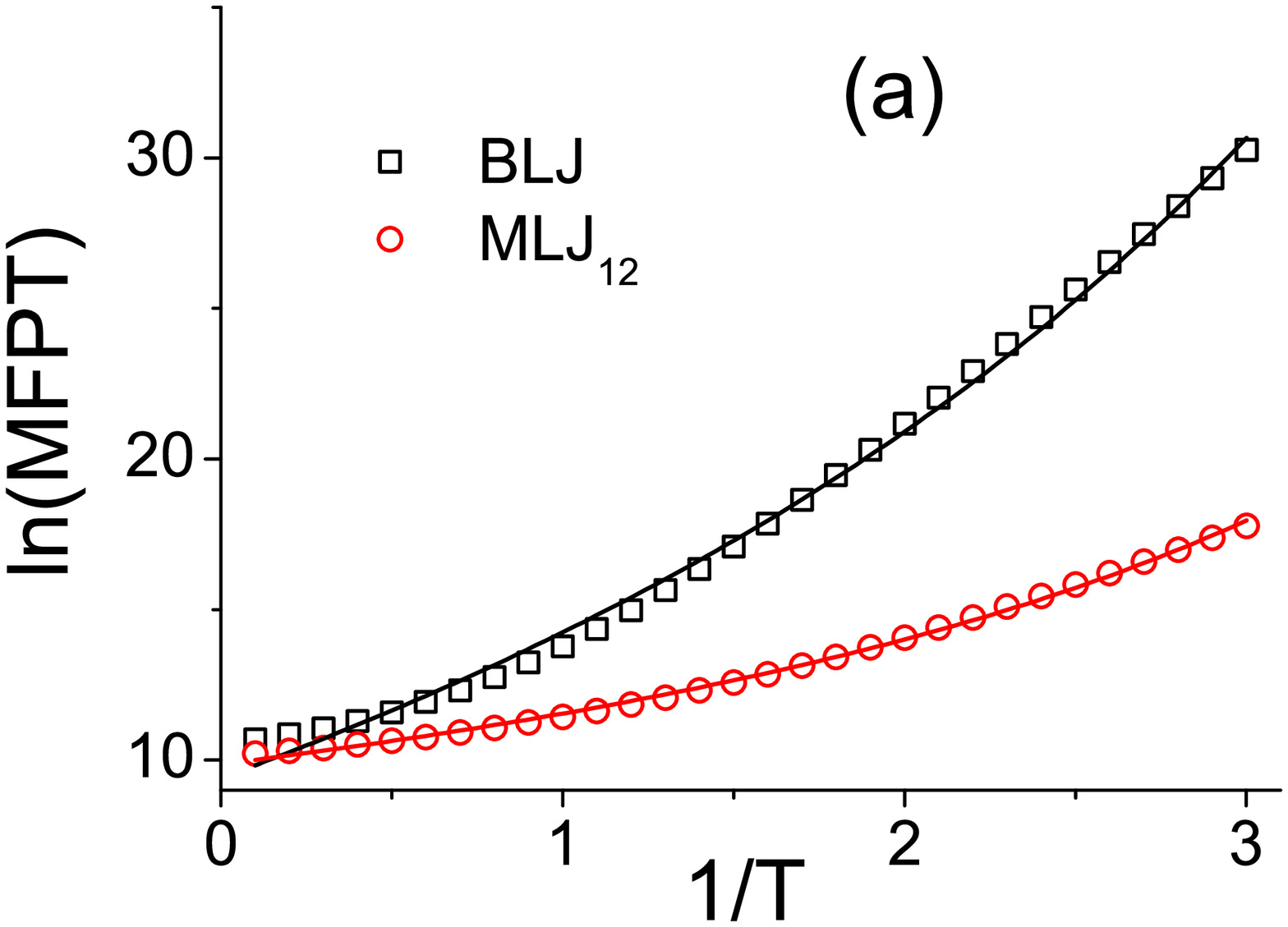}
\includegraphics[width=4cm]{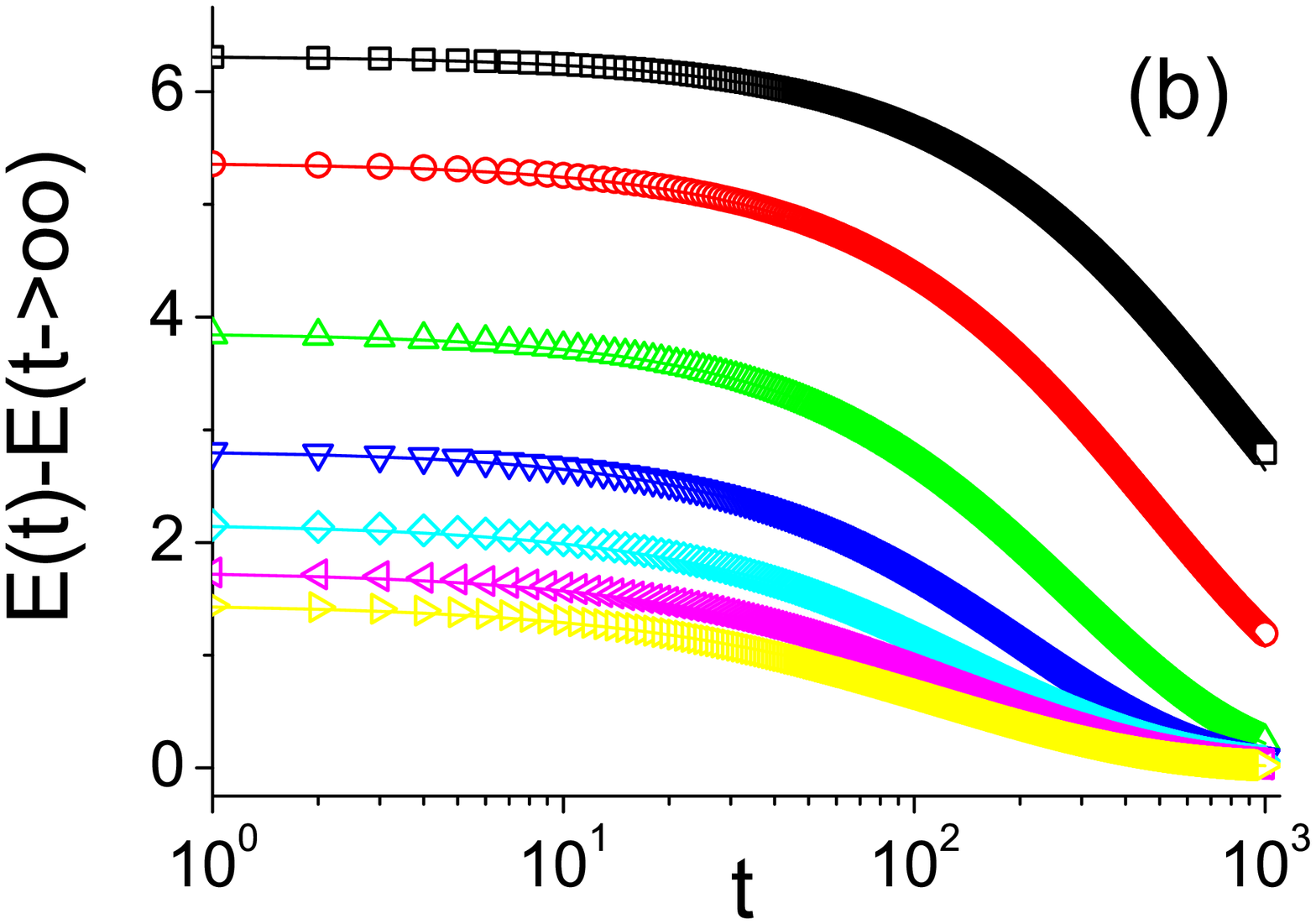} \vspace{-0.5cm}
\end{center} \caption{Dynamical properties of the energy-landscape network.
(a) The mean first passage time (averaged over all sources and
destinations) as a function of the inverse temperature $1/T$, for
MLJ$_{12}$ and BLJ. A super-Arrhenius behavior is observed (the
slope of the curve increases with $1/T$), suggesting that the system
is fragile. The lines are fits to Vogel–-Tammann–-Fulcher law. (b)
For several temperatures (top to bottom: $T=\{2.1,1.8,...,0.3\}$),
the time evolution of the average energy $\overline{E}(T)$ was
calculated. The y-axis shows $\overline{E}(t)-\overline{E}(t
\rightarrow \infty)$ (symbols), such that all curves approach zero.
Curves were fit with a stretched exponential
$\overline{E}(t)-\overline{E}(t \rightarrow \infty) \propto
\exp\left[-(t/\tau)^\beta\right]$ (lines). $\beta \approx 0.8$ and
$\tau$ is between [50,700], and increasing with $1/T$. The picture
is similar for MLJ (not shown).} \label{FPT_correlation}
\end{figure}

Although in our model, microscopic transition rates follow Arrhenius
law, we show below that the global relaxation times deviate from
Arrhenius behavior at low temperatures, suggesting that LJ glass
forming systems are fragile \cite{SupercooledReview}. A global
relaxation time is not naturally defined for the network. However,
we note that as the system evolves in time, it explores the phase
space in a random fashion, according to the transition probabilities
given in Eq. (\ref{transition}). Thus, we associate the global
relaxation time with the time it takes a random walker with
transition probabilities as in Eq. (\ref{transition}) starting at
node $i$, to arrive to node $j$ (the first passage time
\cite{Redner_book}), where $i$ and $j$ are randomly chosen,
uniformly out of all nodes \cite{MFPT_note}. Given the network and
the energy barrier heights, the average first passage time can be
calculated analytically \cite{Redner_book}. In Fig.
\ref{FPT_correlation}(a), we plot the mean first passage time
(averaged over randomly selected sources and destinations) as a
function of the inverse temperature for BLJ and MLJ. A
super-Arrhenius behavior is evident, classifying these systems as a
fragile glass \cite{SupercooledReview}. The data seem to fit to
Vogel–-Tammann–-Fulcher law $\tau\propto\exp\left[A/(T-T_0)\right]$
with $T_0 \approx 0.1$. However, the precise value of $T_0$ highly
depends on the simulation details.

Time dependent quantities can also be studied with the network. For
example, the evolution of the average energy of the system at fixed
temperature can be calculated. We use Eq. (\ref{transition}) and
assume that initially all states are equally probable. The results
are presented in Fig. \ref{FPT_correlation}(b). For short times (up
to about $10^3$ time steps) the decay fits to a stretched
exponential, $\overline{E}(t)-\overline{E}(t \rightarrow \infty) =
A\exp[-(t/\tau)^\beta]$ with $\beta \approx 0.8<1$ \cite{kob}. For
longer times (not shown), the decay is exponential. As in
\cite{AngelaniPRE}, the very fast relaxation, corresponding to
transitions within a basin, is not represented in our model.

\section{Percolation}
\label{Perocaltion_sect}

In addition to dynamical properties, the network topology gives rise
to a static critical temperature $T_p$, where the phase space of
configurations breaks into disconnected components. This is revealed
by percolation theory applied to the energy-landscape network
\cite{percolation_old}. Percolation theory is a powerful framework
for the study of transport in disordered systems. In its simplest
form, it is engaged in the study of conduction in a lattice in which
only a fraction $p$ of the sites, or bonds, are conducting
\cite{bunde_havlin,Stauffer_Book,Kirk_review}. This problem is
relevant in various contexts in which critical phenomena take place,
from superconductors and gelation to forest fires and oil searching.
The theory predicts the value of a critical fraction $p_c$ above
which the bulk sample is conducting, as well as the size, dimension,
total conductance, diffusion coefficients, and other properties of
the percolation clusters.

\begin{figure}[t]
\begin{center}
\includegraphics[width=8cm]{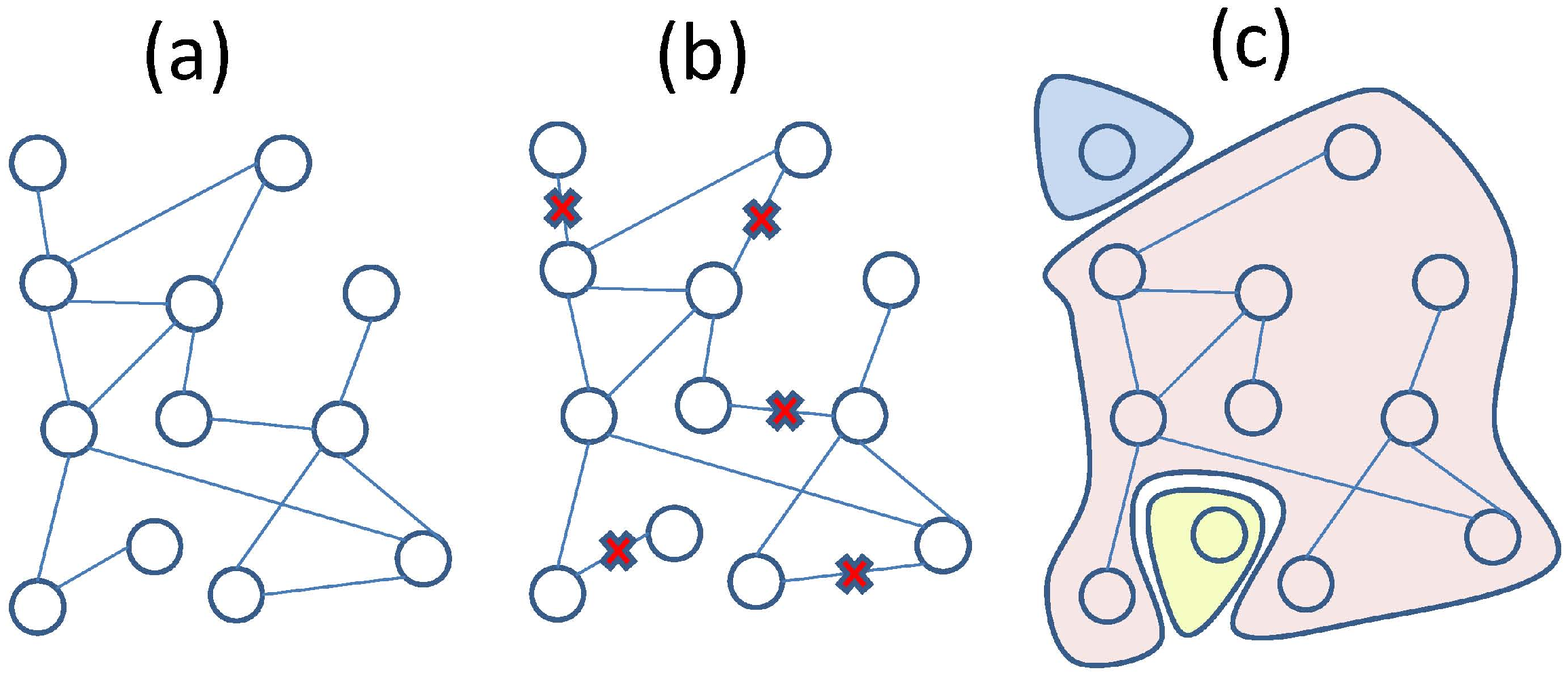} \vspace{-2.3cm}
\end{center} \caption{A schematic of network percolation.
(a) The original network. (b) A fraction $q=1/3$ (5/15) of the links
are removed from the network. (c) The network after removal consists
of one large cluster of 10 nodes and 2 small clusters of one node
each.} \label{schematic_fig}
\end{figure}

In recent years, percolation theory has been successfully applied to
networks to derive criteria for network stability. In a percolation
process over a network, a fraction $q=1-p$ of the network links is
removed \cite{resilience,DM_review}. A percolation transition occurs
when a critical fraction $q_c=1-p_c$ of the links is removed such
that the network disintegrates. The critical point where the network
breaks down is identified by a vanishing size of the largest
connected cluster as well as a divergence in the size of the second
largest cluster \cite{bunde_havlin} (Fig. \ref{schematic_fig}).

At least pictorially, the evolution of the connectivity of the
energy landscape as the temperature is lowered resembles a
percolation process. At high enough temperatures, the system has
sufficient thermal energy to cross most energy barriers. Thus,
connected basins are accessible from each other and the network is
intact. At low temperatures, links which are associated with a
barrier of height $\Delta E \gg T$ can practically not be crossed
and thus can be considered as absent. Thus, as the temperature is
lowered, the network becomes less and less connected, until reaching
the percolation threshold where it fully disintegrates. At that
point, the system is frozen in an isolated region of the landscape,
whose size is a zero fraction of the entire phase space. A
percolation transition of the phase space has been predicted long
ago for spin glasses \cite{percolation_old}. Here, we use the
network representation of LJ clusters to show explicitly how the
percolation transition is realized.

\begin{figure}[t]
\begin{center}
\includegraphics[width=4cm]{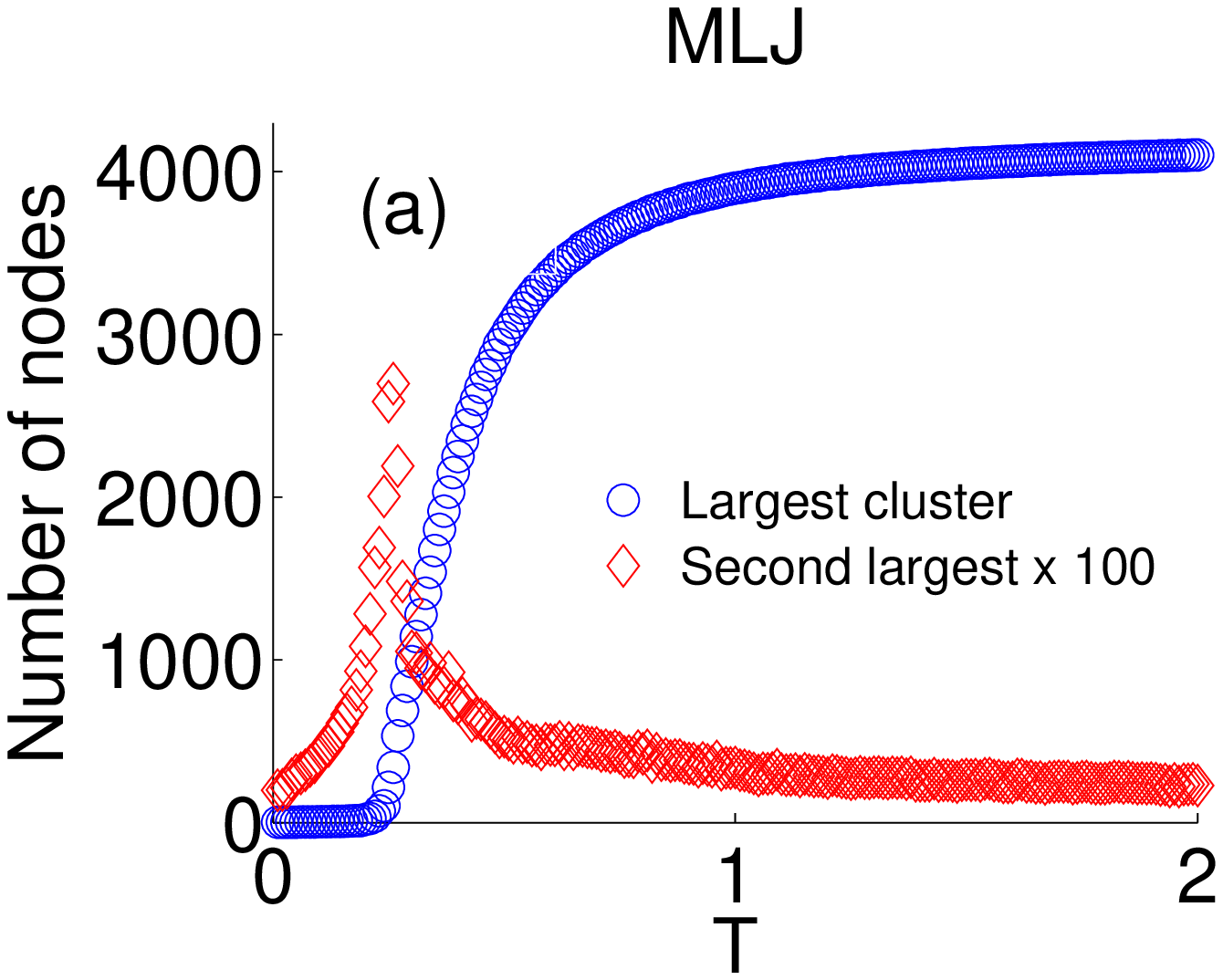}
\includegraphics[width=4cm]{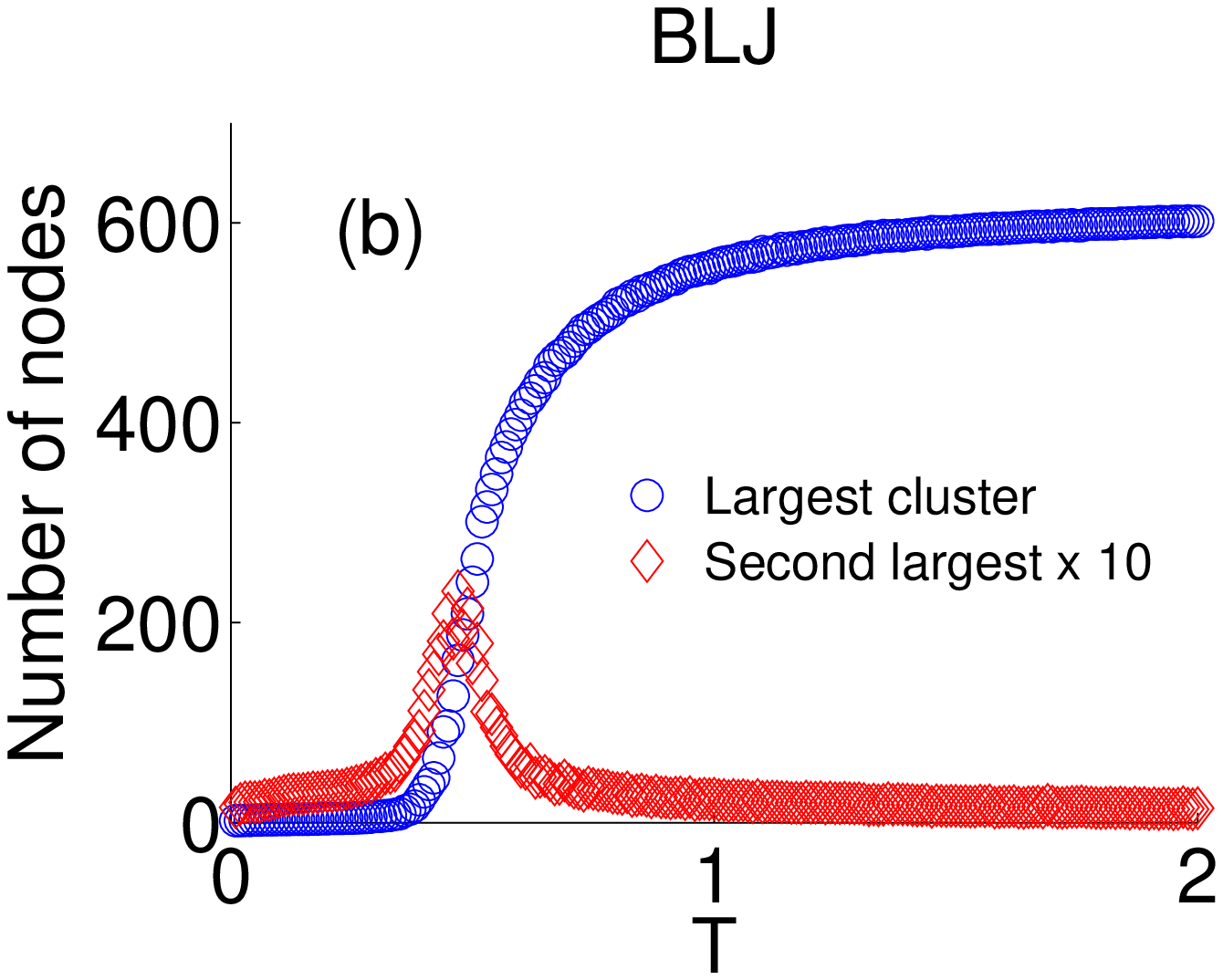} \vspace{-0.5cm}
\end{center} \caption{Percolation transition in Lennard-Jones energy landscape.
(a) For MLJ system, we plot the average size of the largest and
second largest cluster after the removal of each link with
probability $1-e^{-\Delta E/T}$. Clusters are strongly connected
(i.e., each node can be reached from any other node in the cluster).
The percolation transition takes place when the largest cluster size
vanishes and the second largest cluster is maximal. (b) Same as (a)
for BLJ.} \label{percolation_fig}
\end{figure}

Since the probability for a link to be ``active'' decreases with
decreasing temperature, we suggest that links are excluded with
probability $1-e^{-\Delta E/T}$, where $\Delta E$ is the link's
barrier energy. This way, for high $T$, all links remain and the
network is connected, while for low $T$ many links are removed. We
then measure (Fig. \ref{percolation_fig}) the size of the largest
and second largest cluster (where we define a cluster as a set of
nodes mutually accessible from each other) for MLJ and BLJ. A
percolation transition is evident at $T_p=0.26 \pm 0.01$ for MLJ and
$T_p=0.47 \pm 0.01$ for BLJ, indicating a second order phase
transition between a phase where many configurations are available
and a phase with a vanishing number of accessible states.

The percolation transition at $T_p$ is expected to take place at the
final stages of the glass transition, when barriers become almost
impossible to cross, and the system freezes in the glassy state.
Roughly speaking, the percolation transition temperature $T_p$ could
be associated with the Kauzmann temperature $T_K$. At $T_K$, the
configurational entropies of the glass and the crystal are equal
(had the glass transition not intervened), and therefore the system
is bound to a single, non-crystalline, ideal glass state
\cite{SupercooledReview,kob}. Similarly, at $T_p$, the system is
bound to a region of vanishing size of the phase space. In a sense,
this region in phase space corresponds to the ideal glass state in
which the system is found at the Kauzmann temperature $T_K$. However
we emphasize that this correspondence is merely descriptive and
cannot be made more precise.

\section{Discussion}

The understanding of the nature of the glass transition is a
formidable task, particularly since molecular dynamics cannot
approach low enough temperatures. Thus, simplified models which
capture the essential properties of the phenomena are of great
value. Representation of the multidimensional energy surface as a
network is a particularly appealing approach, due to recently
developed network analysis tools. We applied this concept here,
where we studied Lennard-Jones clusters as networks of the stable
basins and the links connecting them, where each link is associated
with an energy barrier. We showed that the network approach
qualitatively reproduces many properties of the glass transition. It
is still not known whether quantitative information, such as the
precise values of $T_g$ and other temperatures can also be extracted
from this kind of analysis. For that purpose, larger systems will
have to be considered. The similarity, in statistical terms, between
the networks of $n=12$ and $n=14$ encourages us to believe that
similar results, at least qualitatively, will be observed in larger
systems.

An alternative approach to circumvent the problem of the small
system size is a mathematical model which captures the main
properties of the energy-landscape network. A naive attempt would be
to construct a ``generalized trap model'' \cite{TrapModel}. In a
regular trap model the configurations of the system are fully
connected, in the sense that the system can jump from any state to
another. However, this is not sufficient to describe the slowing
down of the dynamics, since it allows transitions which do not exist
in reality \cite{Bulbul}. In a trap model adapted to a network, each
configuration is linked to precisely $k$ other configurations, where
it is particularly interesting to consider the case of a power-law
distribution of degrees which is characteristic of LJ (Section
\ref{static}) and other systems \cite{Caflisch}. To complete the
description of the model, one can assume the distribution of energy
barriers is exponential with mean $\Delta E$ (Section \ref{static}).
Despite the attractiveness of this simple approach, our analysis
shows \cite{unpublished} that it leads to counterintuitive results.
For example, the distribution of time $\tau$ the system remains in a
configuration is a power-law $P(\tau) \sim \tau^{-(1+kT/\Delta E)}$.
Thus, according to model, the typical time the system stays at nodes
of high degree is small, whereas it expected that the system will
spend long time at the hubs, since they are found at low potential
energies (Section \ref{static}). Therefore, alternative approaches
should be sought for.

The advantage of the network approach is manifested in the
application of percolation theory, which provides a natural
geometrical interpretation of the structural arrest taking place at
low temperatures \cite{percolation_old}. We studied ``bond
percolation'', where we removed links in which the barrier height
was high relative to the temperature, to reveal a critical
temperature where the phase space breaks down into small isolated
clusters. The study of glassy systems with the network approach can
be further extended. For example, ageing phenomena could be studied,
for either the real network or the model, by introducing more
complex correlation functions. Real-space properties such as
diffusion coefficients and fluctuation-dissipation relations could
be studied by complementing the network with real-space information
for each node. In addition, similar analysis can be pursued to other
systems with complex energy landscapes such as proteins (e.g.,
\cite{Jasna}) or spin glasses \cite{SpinNetwork}.

\begin{acknowledgments}
We thank D. ben-Avraham and S.V. Buldyrev for useful discussions.
Financial support from the National Science Foundation, Israel
Science Foundation, and the Israel Center for Complexity Science is
gratefully acknowledged. S.C. is supported by the Adams Fellowship
Program of the Israel Academy of Sciences and Humanities.
\end{acknowledgments}


\begin{thebibliography}{33}
\bibitem[{KCM()}]{KCM_note}
\bibinfo{note}{Interestingly, kinetically constrained models such as
  facilitated spins \cite{KCM}, show opposite behavior, in which the low energy
  configurations are in many times frozen and have no access to any other
  configuraion.}
\bibitem[{MFP()}]{MFPT_note}
\bibinfo{note}{The relaxation time increases with the inverse temperature in a
  super-Arrhenius form even if the initial sites are weighted by their
  equilibrium occupation probability $\exp[-E_i/T]$.}
\end{thebibliography}

\end{document}